\documentclass[debug]{rmaa}

\usepackage{paralist}
\usepackage{psfrag,color}
\usepackage[latin1]{inputenc}
\usepackage{tabularx}
\usepackage{adjustbox}

\title{Morphological study of a sample of dwarf tidal galaxies using the $C-A$ plane}

\author{ I. Vega-Acevedo,\altaffilmark{1,2,3} 
  and A. M. Hidalgo-G\'amez,\altaffilmark{2}}

\altaffiltext{1}{Departamento de Matem\'aticas, ESIME, Instituto Polit\'ecnico Nacional, M\'exico}

\altaffiltext{2}{Departamento de F\'isica, ESFM, Instituto Polit\'ecnico Nacional, M\'exico}

\altaffiltext{3}{Planetario Luis Enrique Erro, Instituto Polit\'ecnico Nacional, M\'exico.}

\shortauthor{Vega-Acevedo \& Hidalgo-G\'amez}

\shorttitle{TDGs at the $C$-$A$ plane}

\fulladdresses{

\item I. Vega-Acevedo: Escuela Superior de Ingeniere\'ia Mec\'anica y El\'ectrica, Instituto Polit\'enico Nacional, U.P. Adolfo L\'opez Mateos, C.P. 07738, Ciudad de M\'exico, Mexico (ivegaa@ipn.mx).

\item A.M. Hidalgo-G\'amez: Departamento de F\'isica, Escuela Superior de F\'isica y Matem\'aticas, Instituto Polit\'enico Nacional, U.P. Adolfo L\'opez Mateos, C.P. 07738, Ciudad de M\'exico, Mexico (amhidalgog@ipn.mx).}

\listofauthors{I. Vega-Acevedo \& A.M. Hidalgo-G\'amez}

\indexauthor{Vega-Acevedo, I}
\indexauthor{Hidalgo-G\'amez, A.M.}

\abstract{
In this investigation, we determined the Concentration (C) and Asymmetry (A) parameters in a sample of Tidal Dwarf Galaxies (TDG) or candidate galaxies. Most of the galaxies in the sample were found to be in a very precise region of the C-A plane, which clearly separates them from other galaxies. In addition, the stellar mass ($M_{star}$) and the star formation rate ($SFR$) in the sample were determined using optical images and GALEX observations. The main results are: the $M_{star}$ and the $SFR$ in the TDG sample do not follow a linear correlation with the $C$ and $A$ respectively, as observed in the rest of galaxies and the $M_{star}$ and the $SFR$ have a linear correlation similar to that followed by galaxies at high redshift. Then, we can conclude that the C-A plane can be a useful method for the morphological identification of candidates for TDG or dwarf objects from very turbulent environments.}

\resumen{Los par\'ametros morfol\'ogicos Concentraci\'on, $C$, y Asimetr\'ia, $A$, pueden ayudar a identificar si una galaxia enana es candidata a ser una Tidal Dwarf Galaxy (TDG). Se calcularon los valores de $C$ y $A$ en el \'optico de una muestra de galaxias que son TDG o candidatas a serlo. Se encontr\'o que la muestra se identifica f\'acilmente de otros tipos de galaxias. Adem\'as, se determin\'o la masa estelar $M_{star}$, y la formaci\'on estelar, $SFR$, de la muestra empleando im\'agenes en el \'optico y de GALEX. Encontrando que: la $M_{star}$ y $SFR$ no siguen una correlaci\'on lineal respecto a la $C$ y la $A$, tal como se observa para el resto de las galaxias y que la $M_{star}$ y la $SFR$ siguen una correlaci\'on linear similar a la que siguen las galaxias muy corridas al rojo. Por lo tanto, el uso del plano C-A puede ser un m\'etodo para la identificaci\'on morfol\'ogica de candidatos a TDG u objetos enanos de ambientes muy turbulentos.}

\addkeyword{galaxies: dwarf}
\addkeyword{galaxies: irregular}
\addkeyword{galaxies: interactions}
\addkeyword{galaxies: fundamental parameters}
\addkeyword{galaxies: star formation}
\addkeyword{galaxies: structure}

\begin{document}

\maketitle

\section{Introduction}
\label{sec:intro}

A Tidal Dwarf Galaxy (TDG) can be defined as a massive (around $10^{8} M_{\odot}$ baryonic mass) gravitationally bound, self-rotational object of gas, dust, and stars, that is formed during a merger or interaction between massive galaxies \citep{2000AJ....120.1238D}. 
The first time the idea of small galaxies being formed from the debris of interaction between galaxies was proposed by \citet{1978IAUS...77..279S}. Since then, this topic has been very active, identifying these galaxies and studying their main properties e.g. \citep{2012ApJ...750..171S,2010AJ....139.1212S,2004A&A...425..813B,1999IAUS..186...61D,1997A&A...326..537D,1994A&A...289...83D,1992A&A...256L..19M}. These objects typically have an average radius of 6 kpc, an average SFR of $ 8\times 10^{-2}M_{\odot}yr^{-1}$, and a  metallicity of 8.5 dex  \citep{1999IAUS..186...61D}. Some authors, such as \citet{1999IAUS..186...61D}, even mention that TDGs have an average (B-V) color index of 0.3. However, this result is somewhat difficult to establish since \citet{2010AJ....139.1212S} find very dispersed values for the color index (g-r), as well as for the (FUV-g) index, the latter having a very large range of values for all TDGs. Also, they might be relatively long-lived objects, more than 1 Gyr, since after their formation at the interaction, they remain orbiting the parent system or they are expelled from it.  
However, not all of the objects in the tidal tails of the interacting systems are real galaxies. Some are just gas condensations too small to form a gravitationally bounded object (a galaxy) and will dissipate after a few Myr \citep{2004A&A...425..813B}. Others are the result of something called the ``whip effect", which is a phenomenon that occurs when different parts of the tidal tail are superimposed on the line of sight. Using only direct images, it is difficult to differentiate between real Tidal Dwarf Galaxies, gas condensations, and whip-effect objects. Only spectroscopy and, particularly, HI dynamic and position-velocity diagrams, can help to tell them apart. In this sense, the number of real or genuine TDGs is very small, although there are many more tidal objects which cannot be classified as real TDG, but TDG candidates. Also, there are some objects which are in the outermost part of the tidal tails, which have low metallicity and mass. The difficulty here is to determine if these dwarf galaxies are tidal objects or original dwarf galaxies, as in the case of NGC4656 UV \citep{2012ApJ...750..171S}. 
Then, it can be seen that the identification and classification of Tidal Galaxies is a difficult task, and it takes quite a long time to obtain a definitive classification. 

In this investigation, we used the C-A-S system \citep{2003AJ....126.1183C} to separate tidal objects from other types of galaxies, and we checked if it is possible to differentiate between the genuine TDGs candidate TDGs and other tidal objects. The C-A-S system, (Concentration, Asymmetry, and Clumpiness) has recently been used to differentiate morphological types of galaxies \citep{2003ApJS..147....1C}. It has been found that the C-A-S space, or just the $C-A$ plane, is a powerful tool to distinguish between elliptical, spiral, irregular, and starburst galaxies, where each types of galaxy occupies a different region of the C-A-S space \citep{2003ApJS..147....1C,2003AJ....126.1183C}. Moreover, dwarf elliptical galaxies have different values of Concentration and Asymmetry than their larger counterparts \citep{2006MNRAS.368..211Y, 2002AJ....123.2246C, 2003ApJS..147....1C}. Therefore, our goal is to verify if TDGs are located in a separate place in the $C-A$ plane, so they can be easily traced, and if the C and A parameters are different from genuine TDGs as well as for the rest of tidal objects. 

This paper is structured as follows; Section 2 is a description of the sample, the data acquisition, and the determination of parameters. Section 3 presents the result of the $C-A$ plane study for the TDGs and dwarf galaxies. We investigated the possible correlation of $C$ and $A$ with the stellar mass and star formation in  Section 4, along with a discussion of the results, and finally, Conclusions listed in Section 5.


\section{Sample, data acquisition, and determination of the parameters}
\label{subsec:CandA}

As we said, the main goal of this investigation is to study if there are differences in the Concentration, C, and Asymmetry, A, between Tidal Dwarf Galaxies, candidates, and the rest of the dwarf galaxies. In addition, we can check if the C$-$A plane can be used to identify TDGs from other types of tidal objects, like projection effects in the tidal tails.

\subsection{Sample selection}

In order to verify how useful the C$-$A plane is to identify TDGs, we selected a sample of tidal objects,  which includes confirmed TDGs as well as candidates, which are those objects whose self-rotation has not been confirmed yet. We selected 17 objects that come from pre-merger binary pairs with optical tails, merger systems, and interacting galaxies. 

In order to choose the objects of our sample we followed five criteria:  a) only objects classified previously in the literature, as TDGs or TDG candidates, b) objects with a radius strictly less than $6  kpc$, in order to be considered as dwarf galaxies, c) objects in the tidal tails or in the vicinity of galaxies that shown evidence of interactions, d) objects with images in the optical from SDSS and e) with images in the UV by  GALEX. 

The final sample was reduced to a total of $17$ objects, eight of which were confirmed TDGs, five TDG candidates, and four were objects with a very low probability of being TDG. However, there was not enough evidence to affirm the opposite. Due to the criteria imposed for the selection of the sample, the distance, magnitude, as well as other properties of the objects are very different.

In the following, we summarize some of their properties of the selected objects, and some others are listed in Table \ref{tab:sample}. 

The main properties of the sample are listed in Table \ref{tab:sample}. An ID number is given in Column 1, while the name is given in Column 2. This name is that of the parent system along with a letter that describes the location of the TDG. The right ascension and declination are listed in Column 3, while the distance to the system (aka, the interacting parent galaxies) is presented in Column 4. This distance is important, because the further the system is the more difficult it is to distinguish tidal features, and the more easily the low surface brightness structure is lost in the images. In Column 5 are listed the radii of the TDGs, as the r$_{80}$ described in Section 2.2. 

\citet{2010AJ....139.1212S} proposed that Arp 181N, Arp 181S, and Arp 202W could be TDGs. This was confirmed later by \citet{2013MNRAS.431L...1S}. Moreover, \citet{2018MNRAS.475.1148S} found evidence that Arp 202W lacks a significant old stellar population, so they concluded that it might have been formed in the extended dark matter halo of one of its parent galaxies. Also, both systems in Arp 226, Arp 226NW, and Arp 226E, have been confirmed as TDGs by  their metallicity and HI gas dynamics \citep{2015A&A...584A.113L}.
The Arp 244 system was the first where the existence of TDG candidates (Arp 244W and Arp 244E) was reported \citep{1978IAUS...77..279S, 1992A&A...256L..19M}.  The high-resolution mapping of HI made by \citet{2001AJ....122.2969H} and \citet{2001MNRAS.326..578G} corroborated the neutral gas counterpart of these two objects. More recent authors have confirmed that these objects are TDG \citep{2010AJ....139.1212S, 2005ApJ...619L..87H}. Finally, there are two other systems with genuine tidal objects: Arp 245N, which have been proposed as a Tidal Galaxy still in formation \citep{2010AJ....139.1212S, 2001ApSSS.277..405B, 2000AJ....120.1238D}, and Arp 305E, which \citet{2009AJ....137.4643H} proposed as a TDG after studying its star formation and age, all of which was later confirmed by \citet{2017MNRAS.469.3629S}. 

However, there are some other galaxies the nature of which have aroused some doubts. \citet{1994A&A...289...83D},  \citet{1997A&A...326..537D}, and \citet{2010AJ....139.1212S} suggested that Arp 105N and Arp 105S might be tidal objects. However, \citet{2004A&A...425..813B} showed that Arp 105N is a  ``whip effect" object but confirmed that Arp 105S could be a TDG at an early state. Anyhow, we will keep Arp 105N in our sample in order to see if such kind of object can be distinguished from a real TDG in the $C-A$ plane. A similar situation applies for Arp 242, where Arp 242N is considered as a TDG 
\citep{2010AJ....139.1212S}, but \citet{2004A&A...425..813B} did not get any conclusion on Arp 242S. Therefore, it will be considered as a TDG candidate in this investigation. Arp 112E, also called KUG 2359+311, was considered as one of the reddest TDG candidates by \citet{2010AJ....139.1212S}. However, \citet{2020arXiv201106368F} could not observe HI gas in this object, nor any bridge of gas between it and Arp 112. This could indicate two things: it is a normal dwarf galaxy in the vicinity of Arp 112, or KUG 2359+311 is a TDG that has run out of gas. Only \citet{2010AJ....139.1212S} proposed Arp 270N as TDGc, and we will consider it as such in this investigation.  \citet{2008ApJ...676L.113S} found that Holmberg IX is a stronger TDG candidate. Moreover, \citet{2012ApJ...750..171S}, using photometric analysis, also found evidence that Holmberg IX could be a TDGc, although it was not a definitive conclusion. Finally, according to \citet{2012ApJ...750..171S},  NGC 4656 might have, at least, two tidal objects: NGC 4656N and NGC 4656UV. Although  \citet{2017MNRAS.469.4370Z} concluded that NGC4656UV is rather an LSB-dwarf galaxy with dark matter, which agrees with  \citet{2018MNRAS.480.3257M}, where it is proposed that NGC 4656 and 4656UV are a pair of interacting galaxies and NGC 4656UV does not have a tidal origin.

In conclusion, in our sample, there are $8$ confirmed TDG, $4$ candidates, and $5$ objects which might not be TDG, which we call non-likely Tidal Dwarf Galaxy (nlTDG). Although there are a few more TDGs and candidates, these are the ones with good resolution in their optical images, deep enough to allow a good determination of $C$ and $A$. They also have UV images, in order to get the SFR.  Therefore, they are the only ones included in this investigation.

The images used in this work for the determination of C and A were selected from the SDSS in the g filter \citep{2017AJ....154...28B, 2010AJ....139.1628D}, except for Arp 244, where an image from the DSS in the V filter was used. For the determination of the parameters C and A, the MIDAS software was used. The images used for the Star Formation in the Ultraviolet are from the GALEX space observatory database \citep{2017ApJS..230...24B}.

\subsection{Determination of the C and A indexes}

The concentration index, C, quantifies the concentration of the light in a galaxy, and it is defined as 

\begin{equation}
\label{ec.DefofC}
    C=5\times\log\left(\frac{r\left(80\%\right)}{r\left(20\%\right)}\right),
\end{equation}

where r(80\%) and r(20\%) represent the radius that encloses, respectively, 80\% and 20\% of the light curve of the source in units of 1.5 Petrosian reverse radii \citep{2000AJ....119.2645B}. This index has been used extensively to classify galaxies into two broad classes, early and late \citep{2010AJ....139.1628D, 2003ApJS..149..289B}. The correlation between the C index and the stellar mass is very interesting, in the sense that massive galaxies have a higher C index \citep{2006MNRAS.373.1389C}.

 In order to determine $C$, we followed the methodology described below. First, from a two-dimensional Gaussian fitting the optical center of the galaxy was determined. Said point will be the center of elliptical rings, from which the intensity against the radius can be plotted for each object. This can be integrated to get the flux vs. radius plot. Then, we selected those radii which contain 80\% and 20\% of the total flux of the object, respectively. The distance from the center of the galaxy to these points is the r$_{80}$ and r$_{20}$, respectively. These values are the values we used in \ref{ec.DefofC}. Software MIDAS was used. For a more detailed process, the reader should refer to \citet{mastersthesisvega}.

The definition of asymmetry index A we use in this paper is 

\begin{equation}
\label{ec.DefofA}
    A=\frac{\sum\left|I-R\right|}{\sum\left|I\right|},
\end{equation}

 where $I$ is the original image and $R$ is that same image rotated by 180 degrees around the optical center (determined as previously said). The rotated image was created with the software as well as the  $\left|I-R\right|$ one. Both parts of the equation were obtained with the addition of the flux of all the pixels inside the r$_{80}$, and a subsequent division of these two quantities, the total flux in the $\left|I-R\right|$ image and in the $I$ image. The MIDAS software was also used for this procedure \citep{mastersthesisvega}.

The asymmetry index takes values from 0, those galaxies completely symmetrical, to 1 where the galaxy is completely asymmetrical \citep{2003AJ....126.1183C, 2000ApJ...529..886C}. 

The A index has been used to identify recent merger systems that are very distorted. Based on asymmetry measurements on images of nearby merger remnants, it can be considered that a galaxy is a merger remnant if its asymmetry index is larger than a certain value $A>A_{m}$, with $A_{m} = 0.35$ \citep{2003AJ....126.1183C}. Note that this criterion applies to disk-disk mergers only. Spheroid-dominated mergers suffer much weaker morphological distortions, hence this asymmetry criteria cannot be used.

The final values of A and C for this sample of Tidal Dwarf Galaxies are listed in Columns 2 and 3 of Table \ref{tab:CA}, while the stellar mass and the star formation rates are listed in Columns 4 and 5, respectively.

\section{The C and A plane}

Many researchers have used the position in the Concentration$-$Asymmetry plane (C-A) to classify galaxies by their morphology \citep{2000AJ....119.2645B, 2005A&A...434...77L, 2006AJ....131..208M, 2006MNRAS.368..211Y, 2008A&A...478..971H, 2008A&A...484..159N}. Based on  their results we checked if Tidal Dwarf Galaxies were in a separate place in this plane and, therefore, easily spotted.

\begin{figure}
 \includegraphics[width=1\columnwidth]{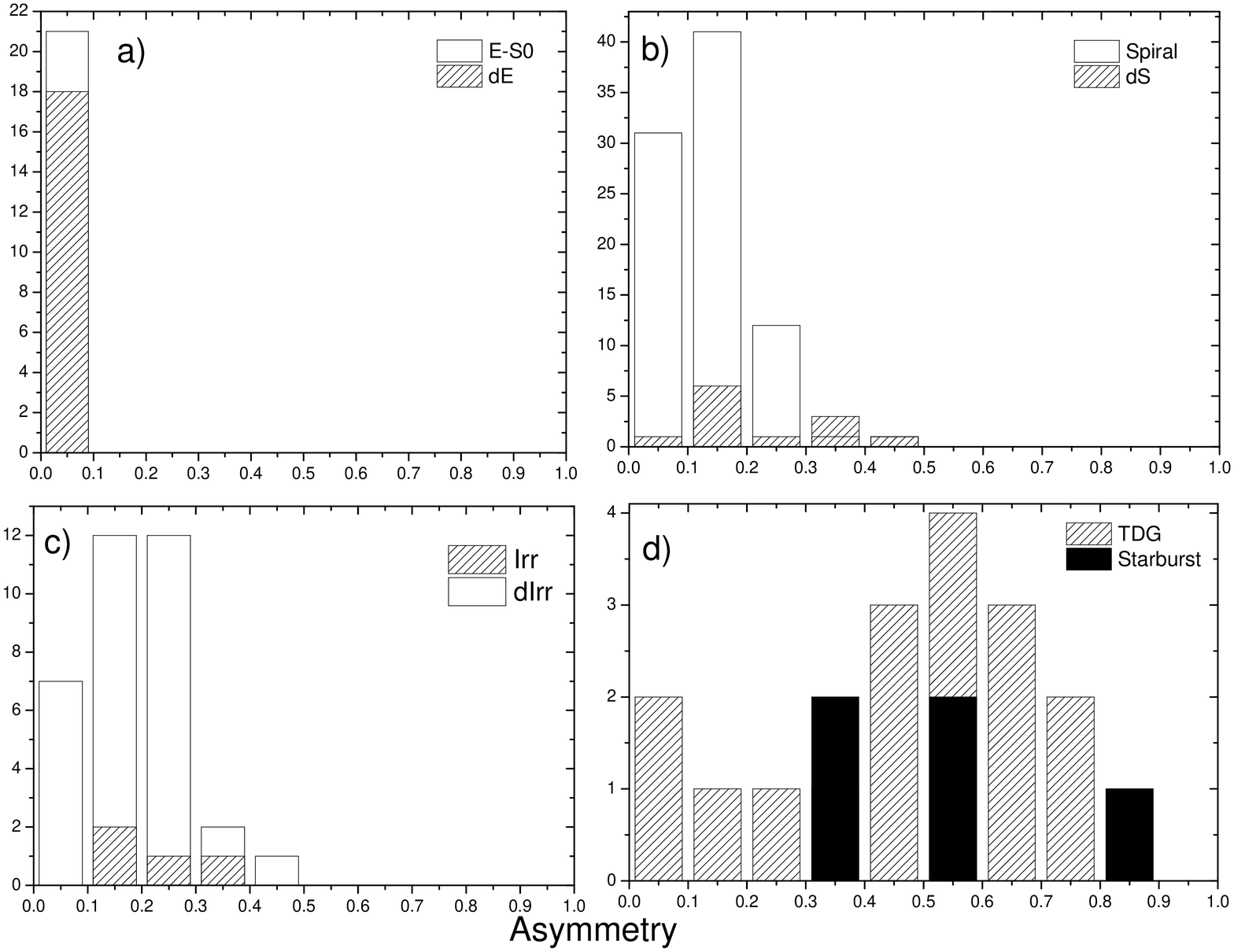}
 \caption{Histogram of asymmetries for \citet{2003ApJS..147....1C} and \citet{2014ASPC..480..239V}}
 \label{fig:HisA}
\end{figure}

The A parameter, listed in Table \ref{tab:CA}, ranges between $0$ and $1$, the smaller the values the more symmetric the galaxy. Only three of our galaxies have low A values ($< 0.2$), but also only three have very high asymmetry values ($> 0.7$). Therefore, most of the TDGs, about 60\% of the galaxies, have intermediate A values. Also, it can be seen that most of the TDG are more asymmetric than the average dwarf galaxies, as can be seen in Table \ref{tab:CA} and Figure \ref{fig:HisA}, where the histogram distribution of the A values for the TDGs in our sample is shown in d) panel. In the other panels the distribution for elliptical a), spiral b) and irregular galaxies c) are shown for comparison. These values have been obtained from \citet{2006MNRAS.373.1389C} and \citet{2014ASPC..480..239V}. There are large differences between TDGs and elliptical  and spiral, the latter having small A values, no larger than $0.3$. On the contrary, irregulars and TDGs show a broad range of asymmetry values, although the peak for Irr (including dwarf) is at lower values than for TDGs.  This is also seen in Table \ref{tab:AveCA}, where the average values of the asymmetry index are shown for some of the galaxies types. TDGs have the largest one, while Irr and dIrr have a smaller average A value (in this investigation we separated between dwarf and normal irregular galaxies). Another interesting aspect to notice is those TDGs with very small asymmetry indexes. As can be seen in Table \ref{tab:CA}, at least two of them have A values lower than $0.1$ (Arp 244S and Arp 245N). We think there are two reasons for such unexpected values: one is that these are not TDGs but some other tidal features more symmetric, which might not be the case based on the large number of investigations that confirmed the tidal nature of these particular objects \citep{2010AJ....139.1212S, 2005ApJ...619L..87H, 2010AJ....139.1212S, 2001ApSSS.277..405B, 2000AJ....120.1238D}. The second one is the lack of low surface brightness structure in the images used for the A determination. This might lead to a lower value of the asymmetry because only the central part of the galaxies was used, which are always more symmetric than the outer parts \citep{mastersthesisvega}. Therefore, it is important to use the deepest images in A determination \citep{mastersthesisvega}. However, as this investigation used archive data, such a requirement could not always be fulfilled.  

Concerning the asymmetry for the different types of tidal objects, the TDGc have the lowest average value ($0.38$) while the nlTDGs have the highest ($0.52$), although the differences are of the same order of the dispersion.

\begin{figure}
 \includegraphics[width=1\columnwidth]{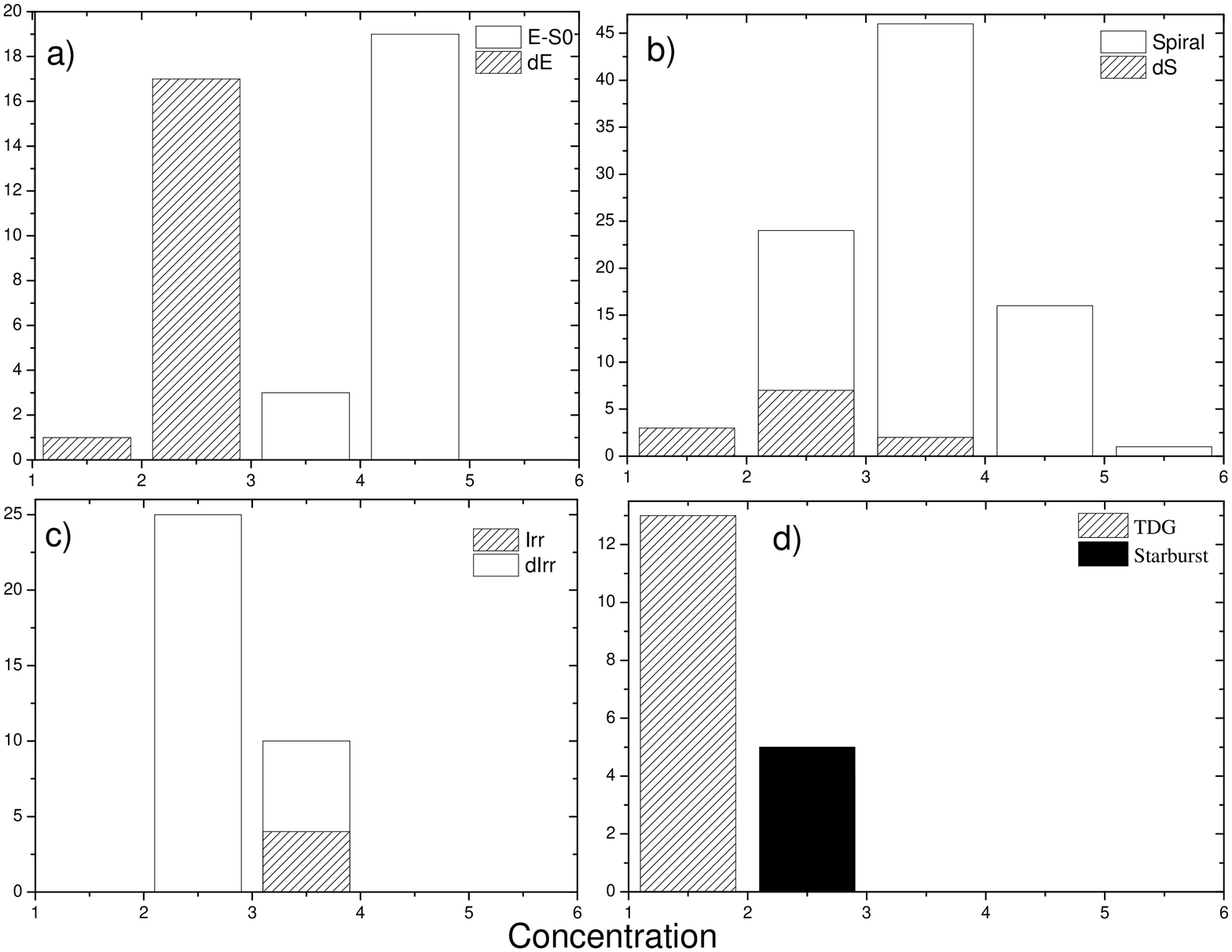}
 \caption{Concentrations histogram for \citet{2003ApJS..147....1C} and \citet{2014ASPC..480..239V}}
 \label{fig:HisC}
\end{figure}

Low values of the C parameter indicate a low concentration of light at the center of the galaxy, which is more common in late-type galaxies (spiral and irregular galaxies). Therefore, \citet{2003ApJS..149..289B} differentiated late and early galaxies based on this parameter. Actually, according to them, a value of C higher than $2.6$ indicates an early galaxy. Only one of the galaxies (Arp 305N) in our sample has C close to this value, while the other $16$ are well into the late-type values. Figure \ref{fig:HisC} shows the histogram distribution of the C values for different morphological types of galaxies. It is clear that TDGs have the narrowest distribution and the lowest C values. Only some of the dS and dE have similar C values, but the average values (listed in column 1 of Table \ref{tab:AveCA}) are very different. The concentration average value is very similar for the three types of dwarf galaxies. Moreover,  it is also interesting to notice that dwarf galaxies always have lower concentration values than their larger counterparts, although all dwarf galaxies have larger C values than the TDGs.  Again, the nlTDG have the highest average concentration values ($1.9$), while the TDGc have the lowest ones ($1.7$). 
In any case, it can be concluded that TDGs have different values of the $A$ and $C$ index than any other types of galaxies, including dwarf ones.

\subsection{$C-A$ plane}

\begin{figure}
 \includegraphics[width=0.6\columnwidth]{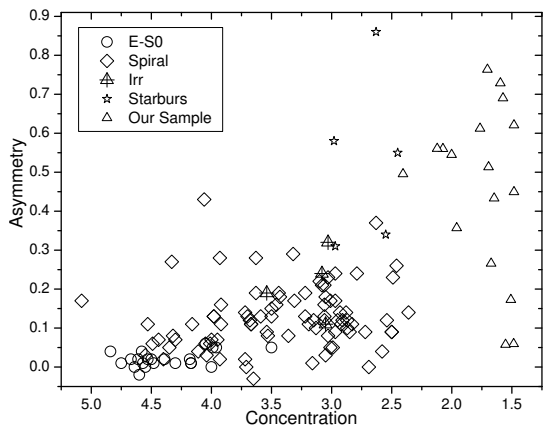}
 \caption{This figure shows the $C-A$ plane for the different morphological types of galaxies. Circles correspond to elliptical or spheroidal galaxies (E, S0), diamonds are for spiral galaxies, cross triangles to irregular galaxies, five-pointed stars to starburst galaxies, and triangles are the objects in our sample. All the data, but that in our sample, were taken from \citet{2003AJ....126.1183C}. \label{fig:AvsC}}
\end{figure}

As seen so far, Tidal Dwarf Galaxies might have particular values of the A and, especially, the C indexes. As pointed out by several investigations \citep{2000ApJ...529..886C, 2003ApJS..147....1C, 2014ASPC..480..239V}, galaxies with different morphologies have different positions in the $C-A$ plane, and they can be differentiated very easily. This is one of the advantages of the CA system. Therefore, we have plotted the TDG of our sample in the $C-A$ plane along with the elliptical, irregular, and spiral galaxies, and Starburst in Figure \ref{fig:AvsC}. There are two interesting conclusions from this figure. Firstly, TDGs are not located near the Starburst galaxies. However, both groups of galaxies have a broad range of A values. Secondly, the C indexes for TDGs are very low. They are the lowest values for all the types of galaxies, despite the low concentration that dwarf galaxies have, as already noted. Therefore, the identification of TDGs can be done very easily using the $C-A$ plane because they are located in a specific strip in this plane, at $C$ values smaller than $2$.

\begin{figure}
 \includegraphics[width=0.6\columnwidth]{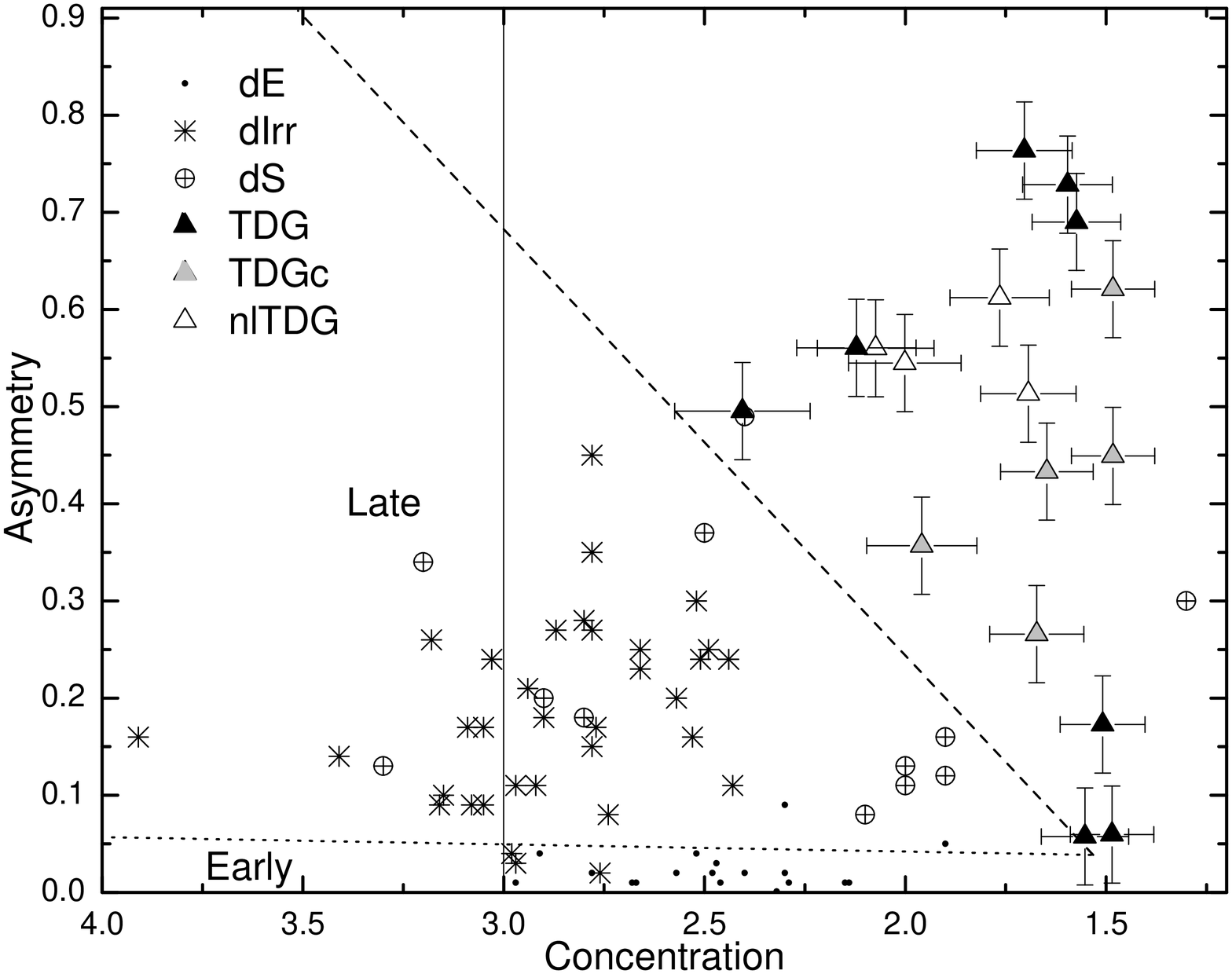}
 \caption{Asymmetry vs Concentration for dwarf galaxies only. This figure shows the $C-A$ plane with different types of dwarf galaxies. Black dots are for dwarf elliptical galaxies (dE), stars correspond to dwarf irregular galaxies (dIrr), dwarf spirals are crossed circles (dS), black triangles are Tidal Dwarf Galaxies confirmed (TDG), grey triangles are Tidal Dwarf Galaxies candidates (TDGc), and white triangles represents the non-likely Tidal Dwarf Galaxies, (nlTDG). The dotted lines separate the early from the late galaxies, and the late from TDGs, while the vertical line represent the maximum value that usually can dwarf galaxies get. The data for dE and dIrr are from \citet{2003AJ....126.1183C}, while the data for dS are from \citet{2014ASPC..480..239V}}.
 \label{fig:AvsCwithDiv}
\end{figure}

In Figure \ref{fig:AvsCwithDiv} we show the $C-A$ plane again, but for dwarf galaxies only. Along with the data points of dE, dS, dI, and TDG, there are the regions of early, and late-type proposed by several authors \citep{2000AJ....119.2645B, 2003AJ....126.1183C}, separated by a dotted line. One of the most striking aspects is that more than 90\% of the dwarf galaxies have a concentration index lower to 3, including dE, but no specific value of A. This is contrary to what happens to large galaxies, where more than $75$\% of them have $C > 3$. Both dS and TDG are located in the late-type region, while dE are located at the bottom of the plane (early region), with very low A indexes and intermediate C values. 
Previous investigations proposed that the structure of dS can be explained by minor interactions \citep{2014ASPC..480..239V}. This might explain the similarities in the C index between dS and TDG's (see Figure \ref{fig:AvsCwithDiv}). The main caveat is the low number of dS galaxies analyzed so far. 
On the contrary, the dIrr galaxies are distributed at large C index ($> 2.5$) and from early to late-type, although almost 70\%  of the dI are in the late-type regions. 

In this figure, it can be seen that TDGs have the same characteristics as other kinds of dwarf galaxies, but they are separated from both dI and dS. They have the smallest C indexes of all the dwarf galaxies, $1.5<C<2.5$, and can be localized in a region over the dashed line in Figure \ref{fig:AvsCwithDiv}, which is given by:

\begin{equation}
    \log\left(A\right)=1.21 C-3.37,
\end{equation}

No other dwarf galaxy is located to the right of this line but one dS. Although more data on late-type dwarf galaxies are needed to reinforce this conclusion, this might indicate the TDGs to be morphologically different from the rest of the dwarf galaxies, with a different origin and evolution.

It is interesting to notice that there is no real difference in the position in the $C-A$ plane between genuine TDG, candidate TDG, and the nlTDG objects.

\section{Discussion}

Although the C parameter values are very similar for most of the galaxies in our sample of Tidal Dwarf Galaxies, the Asymmetry parameter is spread all over the whole range. The relationship between the asymmetry parameter and the Star Formation Rate (Vega-Acevedo \& Hidalgo-G\'amez, in preparation) is well known, so we would like to study the influence of the SFR on the $C-A$ plane, if any. We will also explore how the stellar mass might affect the position of the galaxies in this plane as well.

The Star Formation Rate  for the galaxies in our sample is listed in Table \ref{tab:CA}, Column 4, and it was determined from the UV flux from GALEX with the calibration proposed by  \citet{2010AJ....139..447H}. 

\begin{equation}
    SFR\left[M_{\odot}yr^{-1}\right]=1.27\times10^{-28}L\left[erg s^{-1} Hz^{-1}\right]
\end{equation}

The main caveat is the lack of extinction correction, which could systematically overestimate the SFR. Although it is possible to use IR flux to correct the UV flux from extinction  \citep{2002MNRAS.332..283R}, the small resolution of the infrared images does not allow to obtain the precise IR fluxes of the Tidal Dwarf Galaxies.    

The stellar masses, listed in Table \ref{tab:CA}, Column 3, were determined with the calibrations by \citet{2003ApJS..149..289B}. In particular, we used the following relation

\begin{equation}
    \label{ec.SteMass}
    \log\left[\frac{M_{*}}{L_{g}}\right]=a_{g}+b_{g}\left(g-r\right)
\end{equation}

where the stellar luminosity is given in solar units, and where $a_{g}=-0.499$ and $b_{g}=1.519$ \citep{2003ApJS..149..289B}; and the (g-r) were obtained using the $g$ and $r$ images inside the 
r$_{80}$. From the values listed in Table \ref{tab:CA}, it is clear that most of the  TDG's in our sample have stellar masses lower than  $2\times10^{8}M_{o}$, with a median value of $2.4\times10^8~M_o$, which is very similar to the one determined by \citet{2012MNRAS.419...70K} for a sample of 407 TDGc, of $1.9\times 10^{8}$.

Similarly, the SFR is very low ($<$ 0.1 M$_{\odot}$ yr$^{-1}$) for half of the sample with an average value of $0.17$ M$_{\odot}$ yr$^{-1}$. These values are similar to the typical values in a sample of late-type galaxies \citep{2020RMxAA..56...39M}, although it is very small compared to the SFR of interacting galaxies, which is of the order of $1$-$3.5$ M$_{\odot}$ yr$^{-1}$ \citep{2019A&A...631A..51P}.

\begin{figure}
 \includegraphics[width=0.7\columnwidth]{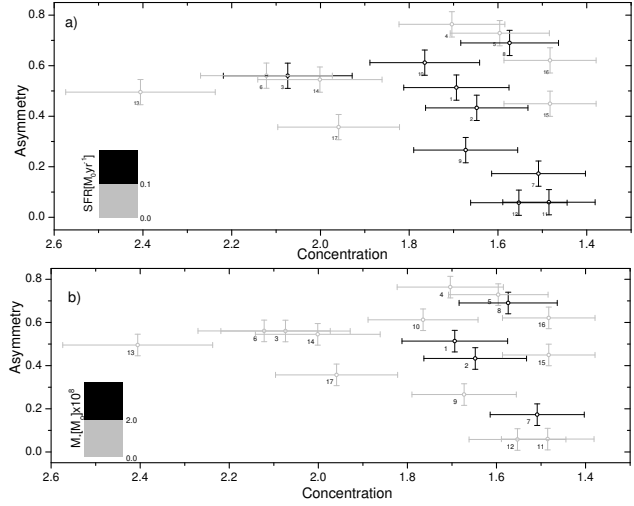}
 \caption{$C-A$ plane. At the top panel, the black points represents objects with SFR over 0.1 M$_{\odot}$ yr$^{-1}$, while grey points represent objects with SFR lower than 0.1 M$_{\odot}$ yr$^{-1}$. At the bottom panel, the black points represent objects with stellar mass larger than $2\times10^{8}M_{\odot}$, and grey points represent those with stellar mass lower $2\times10^{8}M_{\odot}$.}
 \label{fig:AvsCwSMandSFR}
\end{figure}

We can explore again the $C-A$ plane adding these two parameters. They are shown in Figure \ref{fig:AvsCwSMandSFR}, where the different colors indicate different stellar masses at the top panel, and different SFRs at the bottom one. No clear differences can be seen; galaxies with large and small stellar mass are located at the same place in the $C-A$ plane, although all the galaxies with high SFR but one (Arp 112) have $C$ values smaller than $1.8$.

\begin{figure}
 \includegraphics[width=0.6\columnwidth]{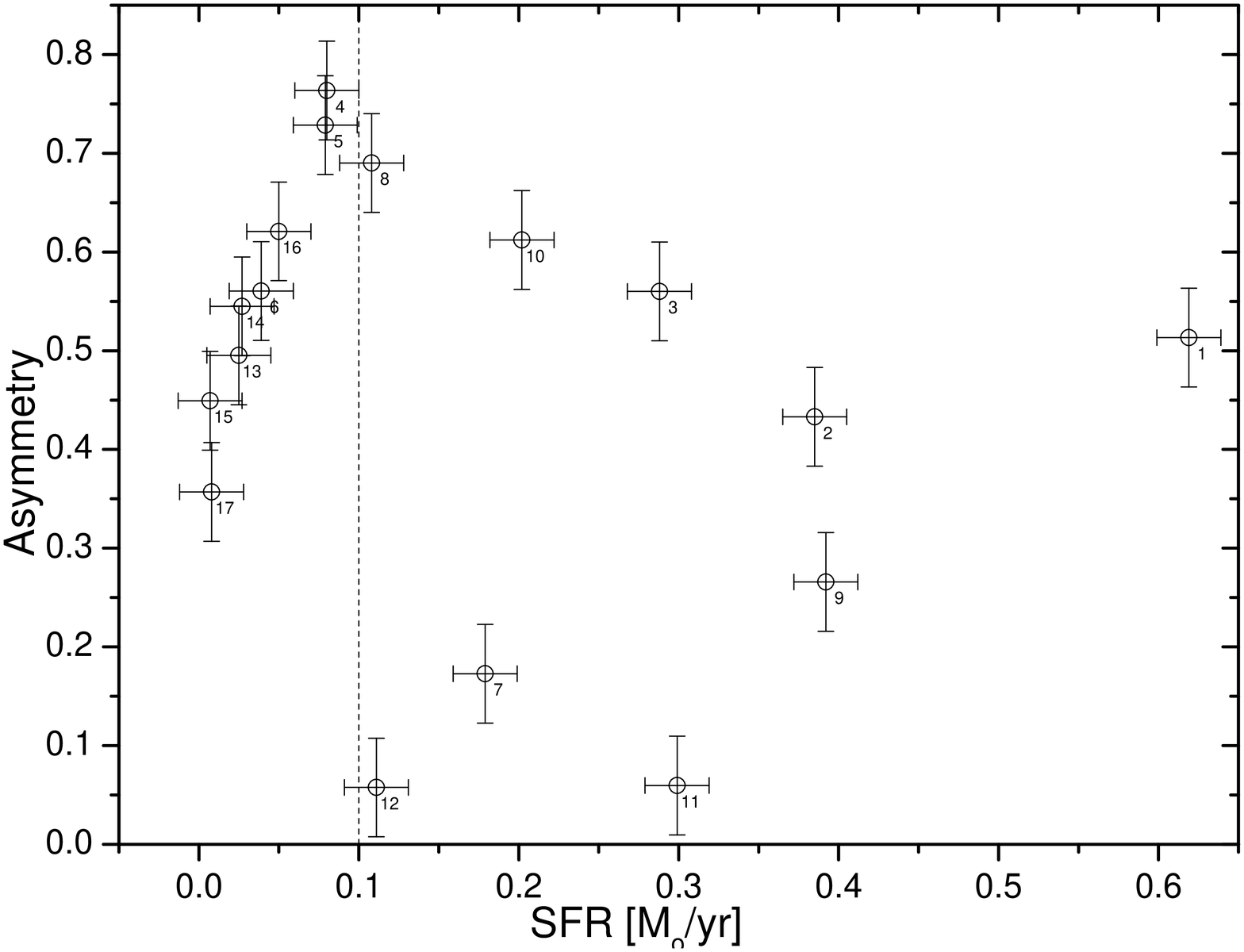}
 \caption{Asymmetry vs the star formation rate. The vertical dash line represents the average value of star formation for dwarf galaxies.}
 \label{fig:AvsSFR}
\end{figure}

In \citet{mastersthesisvega}, as well as in other authors  \citep{2003AJ....126.1183C, 2001RMxAC..11..115M}, a relationship between the asymmetry parameter and the SFR was obtained for normal galaxies. We can see that the TDG in Figure \ref{fig:AvsSFR} can be grouped into two categories: those with a SFR lower than $0.1$~M$_{\odot}$ yr$^{-1}$ seem to follow a linear correlation, while TDGs with a higher SFR do not show a clear correlation, only a dispersion diagram, with no particular value of the asymmetry. Moreover, there are no real differences between the candidates, the confirmed TDGs, and the nlTDG in the diagram, although three out of five of the latter are located at the high SFR locus. More data are needed to confirm the lack of trend for galaxies with high SFR. We must notice that the SFR determined in \citet{2014ASPC..480..239V} was using the H$_{\alpha}$ flux, which gives smaller values than the $UV$ flux, so small differences are expected.

\begin{figure}
 \includegraphics[width=0.6\columnwidth]{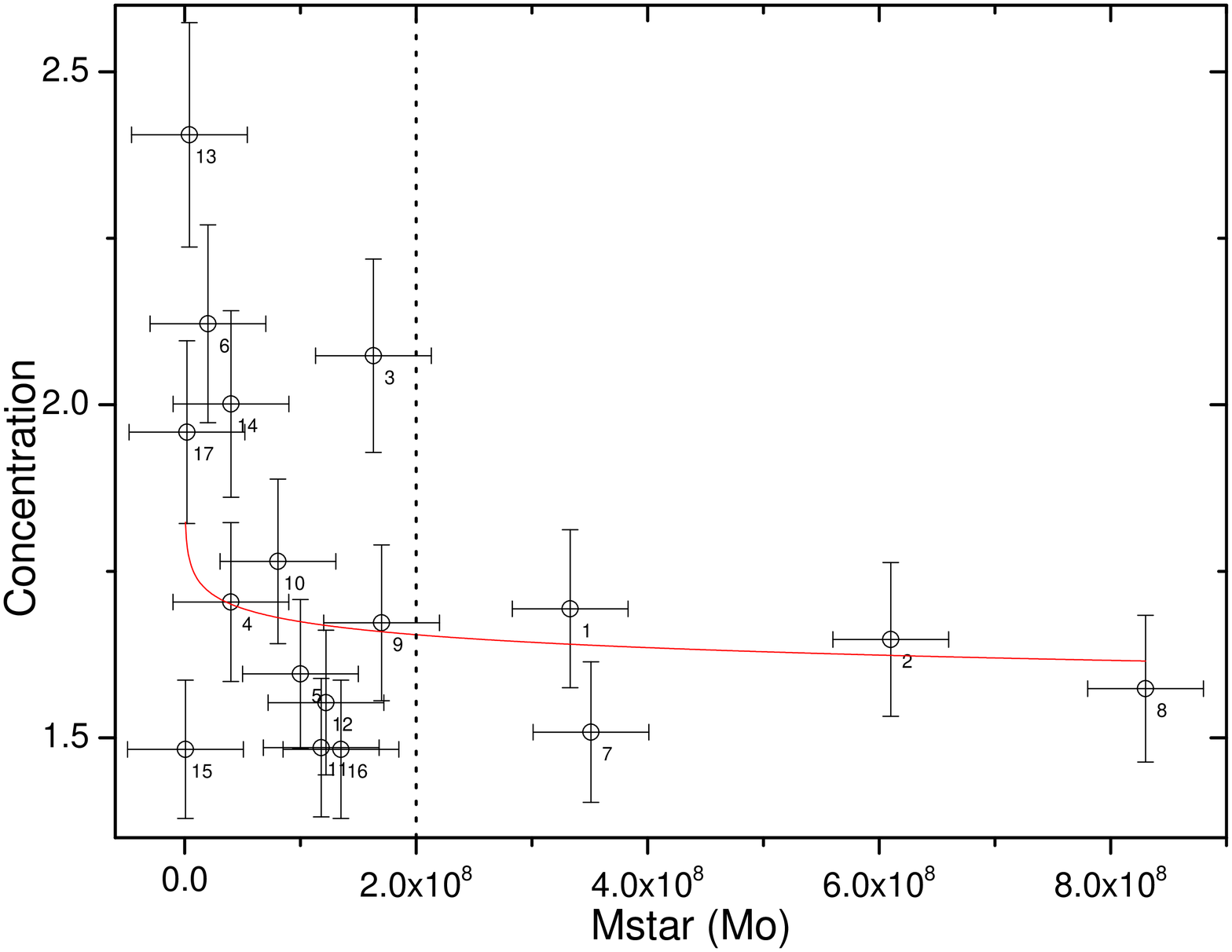}
 \caption{Concentration vs stellar mass, $M_{star}$. The vertical dash line represents the average value of stellar mass for dwarf galaxies, while the solid line is fitting to the data-points.}
 \label{fig:MsvsC}
\end{figure}

We can also check if there is any relationship between C and the stellar mass as  proposed for normal galaxies \citep{2006MNRAS.373.1389C}. Figure \ref{fig:MsvsC} shows that there is a logarithmic relationship, 

\begin{equation}
    \label{ec.CrelSteMass}
    C=ax^{b},
\end{equation}

where $a=2.28\pm 0.68$ and $b=-0.01\pm 0.01$ (see solid line in Figure \ref{fig:MsvsC}). Despite this possible correlation, it is clear to see that the TDG can be divided into two groups, with approximately 76\% of the sample having a stellar-mass of less than $2\times 10^{8}M_{\odot}$.

\begin{figure}
 \includegraphics[width=0.6\columnwidth]{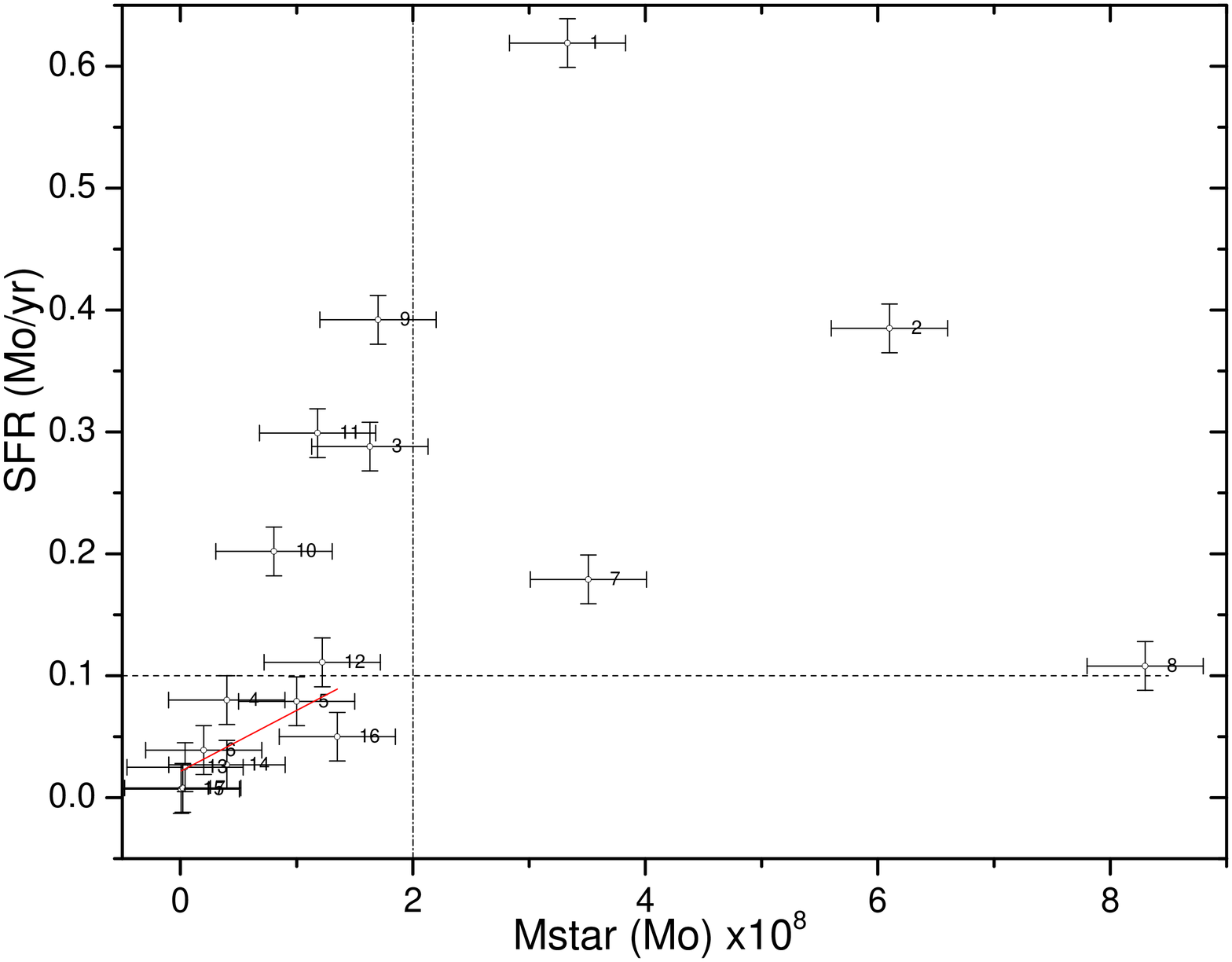}
 \caption{Star formation rate, $SFR$ vs stellar mass, $M_{star}$. The horizontal dash line represents the average value of star formation for dwarf galaxies, and  the vertical dash line represents the average value of stellar mass for dwarf galaxies. The solid line is fitting to the data-points. }
 \label{fig:SFRvsMs}
\end{figure}

In a recent study of galaxies in chaotic environments at high redshift (Cosmic Assembly Near-infrared Deep Extragalactic Legacy Survey, CANDELS) by \citet{2015ApJ...799..183S}, a linear correlation between the SFR and the stellar mass was found. Although CANDELS galaxies are early-type at $z$ between $4$ and $6$, it is assumed that its gas is very turbulent. In such a way, they are very similar to TDGs, which are being formed from a very turbulent environment. Therefore, a similar correlation might be expected for the latter.  As can be seen in Figure \ref{fig:SFRvsMs}, there is a linear correlation 

\begin{equation}
    \label{ec.SFRreSM}
SFR~(M_{\odot}yr^{-1})=\alpha~M_{*}~(M_{o})+\beta,    
\end{equation}

where $\alpha=(5\pm1)\times10^{-10}$ and $\beta=0.02\pm0.01$. These values are very similar to those of CANDELS galaxies for $z=4$. It is very interesting to note that those galaxies with a stellar mass of less than $2 M_{\odot}$ and a stellar formation of less than $0.1~M_{o}yr^{-1}$ are those reinforcing such relationship, while only two galaxies outside this ``box" follow it. Galaxies with low $M_{*}$ but high SFR seems to follow a linear correlation but with a different slope.  More data are needed to understand which is the reason for these galaxies to not follow a single correlation between the SFR and the stellar mass.

\section{Conclusions}
\label{sec:Concl}

In this investigation, we determined the $C$ and $A$ parameters for a sample of TDGs in order to know if they have similar values to the rest of the galaxies. In our sample, we included five tidal objects, which might not be TDGs just to check if there are any differences between genuine, real TDGs and any other tidal object. 

We noticed that the TDGs have the lowest $C$ values than any other group of galaxies. This indicates that they are the loosest of the galaxies, which is quite expected if these objects are still in the assembling stage. Moreover, the exact value of the $C$ parameters does not depend on the stellar mass or the SFR. Also, TDG are the galaxies with the largest values of the $A$ parameter, except for Starburst galaxies. This is quite expected because they are just assembling and the asymmetry might be larger. 

With these values, it is clear that TDGs are located in a separated, well defined region in the $C-A$ plane. No other galaxies, including dwarf galaxies, are in this part of the plane. This might indicate that TDGs have a very different origin than the rest of the galaxies. Or the other way around, dwarf galaxies (dE, dS, and dI) are not formed from the debris of interacting systems. The main caveat is that there is no difference between the confirmed TDGs, the TDG candidates, and the non-likely TDG objects with very similar average values of both $A$ and $C$ for the three subsamples. 

Contrary to what was obtained in large galaxies, there is no correlation between $A$ parameter and SFR, except for galaxies with SFR smaller than $0.1$ M$_{\odot}$ yr$^{-1}$. Moreover, the relationship between $C$ and stellar mass is exponential but with large dispersion. 

On the other hand, we obtained a relationship between SFR and the stellar mass for the galaxies in our sample. This correlation is very similar to the one obtained for CANDELS galaxies, which are early-type galaxies at $z>4$ with a turbulent environment. This is interesting because TDGs, as they are being formed during an interaction of galaxies, also have a very turbulent environment. This suggests that the environment and turbulence are crucial parameters for understanding galaxy formation. However, for a thorough conclusion on this subject, more TDGs are needed.

\section{Acknowledgements}
\addcontentsline{toc}{section}{Acknowledgements}
 This research was supported by the Instituto Polit\'ecnico Nacional (M\'exico) under the research projects  SIP-20210556, and SIP-20200822. This work is part of M. en C. I. Vega Acevedo's Ph.D. thesis, sponsored by CONACyT. This investigation has made use of the NASA/IPAC Extragalactic Database (NED), which is operated by the Jet Propulsion Laboratory, California Institute of Technology, under contract with the National Aeronautics and Space Administration (NASA). This work is based in part on observations made with the GALEX Space Telescope, which is operated by the Jet Propulsion Laboratory with the National Aeronautics and Space Administration (NASA). The SDSS-III web site is http://www.sdss3.org/. SDSS-III is managed by the Astrophysical Research Consortium for the Participating Institutions of the SDSS-III Collaboration including the University of Arizona, the Brazilian Participation Group, Brookhaven National Laboratory, Carnegie Mellon University, University of Florida, the French Participation Group, the German Participation Group, Harvard University, the Instituto de Astrofisica de Canarias, the Michigan State/Notre Dame/JINA Participation Group, Johns Hopkins University, Lawrence Berkeley National Laboratory, Max Planck Institute for Astrophysics, Max Planck Institute for Extraterrestrial Physics, New Mexico State University, New York University, Ohio State University, Pennsylvania State University, University of Portsmouth, Princeton University, the Spanish Participation Group, University of Tokyo, University of Utah, Vanderbilt University, University of Virginia, University of Washington, and Yale University. The authors wish to thank Mr. P. G{\'o}mez-Garc{\'e}s and Lic. A. Romo-Gonz{\'a}lez for a careful revision of the English grammar.

\begin{table}[!t]\centering
  \setlength{\tabnotewidth}{1\columnwidth}
  \tablecols{5}
  \setlength{\tabcolsep}{2.3\tabcolsep}
  \caption{Sample of Tidal Dwarf Galaxies.} \label{tab:sample}
 \begin{tabular}{clccc}
    \toprule
    ID$^{a}$ (1) & Name (2) & RA Dec $\left[J2000\right]$ (3) & Dis$\left[Mpc\right]$ (4) & Radii$^{b}\left[kpc\right]$ (5)\\
    \midrule
1	&	Arp 105N$^{3}$	&	11:11:12.8 28:45:57.14	&	134.25	&	4.71\\
2	&	Arp 105S$^{2}$	&	11:11:13.4 28:41:15.96	&	134.25	&	4.71	\\
3	&	Arp 112E$^{3}$	&	00:01:34.5 31:26:33.70	&	66.09	&	5.31	\\
4	&	Arp 181W$^{1}$	&	10:27:26.3 79:49:12.79	&	143.56	&	5.30	\\
5	&	Arp 181E$^{1}$	&	10:27:40.1 79:49:45.3	&	143.56	&	3.88  	\\
6	&	Arp 202W$^{1}$	&	09:00:09.3 35:43:40.26	&	48.70	&	2.63  	\\
7	&	Arp 226NW$^{1}$	&	22:20:33.5 -24:37:22.07	&	66.14	&	3.85  	\\
8	&	Arp 226E$^{1}$	&	22:20:55.7 -24:41:10.21	&	66.14	&	4.47 	\\
9	&	Arp 242N$^{2}$	&	12:46:10.4 30:45:11.31	&	101.44	&	4.15  	\\
10	&	Arp242S$^{3}$	&	12:46:12.0 30:42:02.34	&	101.44	&	5.62  	\\
11	&	Arp244S$^{1}$	&	12:01:26.6 -19:00:49.33	&	30.38	&	3.95 	\\
12	&	Arp245N$^{1}$	&	09:45:44.1 -14:17:34.55	&	39.15	&	5.81 	\\
13	&	Arp305E$^{1}$	&	11:58:41.5 27:29:34.90	&	55.12	&	4.86  	\\
14	&	NGC4656N$^{3}$	&	12:44:14.4 32:16:43.88	&	13.41	&	5.40	\\
15	&	Arp270S$^{2}$	&	10:49:34.5 32:52:38.31	&	28.14	&	1.91	\\
16	&	Arp270N$^{2}$	&	10:49:44.2 33:00:42.40	&	28.14	&	0.76 	\\
17	&	HolmbergIX$^{2}$	&	09:57:31.5 69:02:43.69	&	1.90	&	0.86    \\
    \midrule
    \bottomrule
    \tabnotetext{a}{Column (1) is the Identification number in each object of our sample.}
    \tabnotetext{b}{The radii in column (6) are the r(80\%),  obtained as described in Section 2.2.}
    \tabnotetext{1}{Tidal Dwarf Galaxies confirmed, TDG.}
    \tabnotetext{2}{Tidal Dwarf Galaxies candidates, TDGc.}
    \tabnotetext{3}{non-likely Tidal Dwarf Galaxies, nlTDG.}
  \end{tabular}
\end{table}

\begin{table}[!t]\centering
  \setlength{\tabnotewidth}{0.5\columnwidth}
  \tablecols{3}
  \setlength{\tabcolsep}{2.8\tabcolsep}
  \caption{Sample values } \label{tab:CA}
 \begin{tabular}{lcccc}
    \toprule
    ID$^{a}$ & \multicolumn{1}{c}{A} & \multicolumn{1}{c}{C} & \multicolumn{1}{c}{$M_{*}\left[M_{o}\right]10^{8}$} & \multicolumn{1}{c}{SFR$\left[M_{o}yr^{-1}\right]10^{-2}$}  \\
    \midrule
1	&	0.5	$\pm$	0.1	&	1.7	$\pm$	0.3	&	3.3 $\pm$	0.5	&	61.9 $\pm$	0.5	\\
2	&	0.4	$\pm$	0.1	&	1.7	$\pm$	0.3	&	6.1 $\pm$	0.5	&	38.5 $\pm$	0.5	\\
3	&	0.6	$\pm$	0.1	&	2.1	$\pm$	0.3	&	1.6 $\pm$	0.5	&	28.8 $\pm$	0.5	\\
4	&	0.8	$\pm$	0.1	&	1.7	$\pm$	0.3	&	0.4 $\pm$ 0.5	&	8.0 $\pm$	0.5	\\
5	&	0.7	$\pm$	0.1	&	1.6	$\pm$	0.2	&	1.0 $\pm$ 0.5	&	7.9 $\pm$	0.5	\\
6	&	0.6	$\pm$	0.1	&	2.1	$\pm$	0.3	&	0.2 $\pm$ 0.5	&	3.9 $\pm$	0.5	\\
7	&	0.2	$\pm$	0.1	&	1.5	$\pm$	0.2	&	3.5 $\pm$ 0.5	&	17.9 $\pm$	0.5	\\
8	&	0.7	$\pm$	0.1	&	1.6	$\pm$	0.2	&	8.3 $\pm$ 0.5	&	10.8 $\pm$	0.5	\\
9	&	0.3	$\pm$	0.1	&	1.7	$\pm$	0.3	&	1.7 $\pm$ 0.5	&	39.2 $\pm$	0.5	\\
10	&	0.6	$\pm$	0.1	&	1.8	$\pm$	0.3	&	0.8 $\pm$ 0.5	&	20.2 $\pm$	0.5	\\
11	&	0.1	$\pm$	0.1	&	1.5	$\pm$	0.2	&	1.2 $\pm$ 0.5	&	29.9 $\pm$	0.5	\\
12	&	0.1	$\pm$	0.1	&	1.6	$\pm$	0.2	&	1.2 $\pm$ 0.5	&	11.1 $\pm$	0.5	\\
13$^{a}$	&	0.5	$\pm$	0.1	&	2.4	$\pm$	0.4	&	--	&	2.5 $\pm$	0.5	\\
14	&	0.5	$\pm$	0.1	&	2.0	$\pm$	0.3	&	0.40 $\pm$ 0.5	&	2.7 $\pm$	0.5	\\
15$^{a}$	&	0.5	$\pm$	0.1	&	1.5	$\pm$	0.2	&	--	&	0.7 $\pm$	0.5	\\
16	&	0.6	$\pm$	0.1	&	1.9	$\pm$	0.3	&	1.4 $\pm$ 0.5	&	5.0 $\pm$	0.5	\\
17$^{a}$	&	0.4	$\pm$	0.1	&	2.0	$\pm$	0.3	&	--	&	0.8 $\pm$	0.5	\\

    \midrule
   
    \bottomrule
    \tabnotetext{a}{The stellar mass obtained for these objects is lower than $0.1\times10^{8}M_{\odot}$.}
  \end{tabular}
\end{table}

\begin{table}[!t]\centering
  \setlength{\tabnotewidth}{1\columnwidth}
  \tablecols{3}
  \setlength{\tabcolsep}{2.8\tabcolsep}
  \caption{Averages and 1 $\sigma$ Variations of C and A for Galaxy Types} \label{tab:AveCA}
 \begin{tabular}{lcccc}
    \toprule
    Type & \multicolumn{1}{c}{C} & \multicolumn{1}{c}{A} \\
    \midrule
TDGs	&	1.7	$\pm$	0.2	&   0.5	$\pm$	0.2	  \\
Ellipticals$^{a}$	&	4.4	$\pm$	0.3	&	0.0	$\pm$	0.1  \\
Dwarf ellipticals$^{a}$	&	2.5	$\pm$	0.3	&	0.0	$\pm$	0.1  \\
Spiral$^{a}$ & 3.3 $\pm$ 0.6 & 0.1 $\pm$ 0.1 \\
Dwarf Spiral$^{b}$ & 2.4 $\pm$ 0.6 & 0.2 $\pm$ 0.1 \\
Irregulars$^{a}$	&	3.3	$\pm$	0.5	&	0.3  $\pm$	0.2  \\
Dwarf irregulars$^{a}$ &	2.9	$\pm$	0.3	&	0.2  $\pm$	0.1	  \\
    \midrule
   
    \bottomrule
    \tabnotetext{a}{Data taken from \citet{2003ApJS..147....1C}}
    \tabnotetext{b}{Data taken from \citet{2014ASPC..480..239V}}
    \tabnotetext{}{The error for C and A corresponds to a 1 $\sigma$ variation from the average.}
  \end{tabular}
\end{table}


\begin{thebibliography}

\bibitem[\protect\citeauthoryear{Bell et al.}{2003}]{2003ApJS..149..289B} Bell E.~F., McIntosh D.~H., Katz N., Weinberg M.~D., 2003, ApJS, 149, 289

\bibitem[\protect\citeauthoryear{Bershady et al.}{2000}]{2000AJ....119.2645B} Bershady M.~A., Jangren A., Conselice C.~J., 2000, AJ, 119, 2645

\bibitem[\protect\citeauthoryear{Bianchi, Shiao, \& Thilker}{2017}]{2017ApJS..230...24B} Bianchi L., Shiao B., Thilker D., 2017, ApJS, 230, 24. doi:10.3847/1538-4365/aa7053

\bibitem[\protect\citeauthoryear{Blanton et al.}{2017}]{2017AJ....154...28B} Blanton M.~R., Bershady M.~A., Abolfathi B., Albareti F.~D., Allende Prieto C., Almeida A., Alonso-Garc{\'\i}a J., et al., 2017, AJ, 154, 28. doi:10.3847/1538-3881/aa7567

\bibitem[\protect\citeauthoryear{Bournaud et al.}{2004}]{2004A&A...425..813B} Bournaud F., Duc P.-A., Amram P., Combes F., Gach J.-L., 2004, A\&A, 425, 813. doi:10.1051/0004-6361:20040394

\bibitem[\protect\citeauthoryear{Brinks et al.}{2001}]{2001ApSSS.277..405B} Brinks E., Duc P.-A., Springel V., Pichardo B., Weilbacher P., Mirabel F., 2001, ApSSS, 277, 405. doi:10.1023/A:1012719910476


\bibitem[\protect\citeauthoryear{Conselice}{2003}]{2003ApJS..147....1C} Conselice C.~J., 2003, ApJS, 147, 1

\bibitem[Conselice (2006a)]{2006MNRAS.373.1389C} Conselice C.~J., 2006, MNRAS, 373, 1389

\bibitem[\protect\citeauthoryear{Conselice, Bershady \& Gallagher}{2000}]{2000A&A...354L..21C} Conselice C.~J., Bershady M.~A., Gallagher J.~S., 2000, A\&A, 354, L21

\bibitem[\protect\citeauthoryear{Conselice et al.}{2000}]{2000ApJ...529..886C} Conselice C.~J., Bershady M.~A., Jangren A., 2000, ApJ, 529, 886

\bibitem[\protect\citeauthoryear{Conselice et al.}{2003}]{2003AJ....126.1183C} Conselice C.~J., Bershady M.~A., Dickinson M., Papovich C., 2003, AJ, 126, 1183


\bibitem[\protect\citeauthoryear{Conselice et al.}{2000}]{2000AJ....119...79C} Conselice C.~J., Gallagher J.~S., Calzetti D., Homeier N., Kinney A., 2000, AJ, 119, 79

\bibitem[\protect\citeauthoryear{Conselice et al.}{2002}]{2002AJ....123.2246C} Conselice C.~J., Gallagher J.~S., Wyse R.~F.~G., 2002, AJ, 123, 2246. doi:10.1086/340081

\bibitem[\protect\citeauthoryear{Conselice, Rajgor \& Myers}{2008}]{2008MNRAS.386..909C} Conselice C.~J., Rajgor S., Myers R., 2008, MNRAS, 386, 909

\bibitem[\protect\citeauthoryear{Doi et al.}{2010}]{2010AJ....139.1628D} Doi M., Tanaka M., Fukugita M., Gunn J.~E., Yasuda N., Ivezi{\'c} {\v{Z}}., Brinkmann J., et al., 2010, AJ, 139, 1628. doi:10.1088/0004-6256/139/4/1628


\bibitem[\protect\citeauthoryear{Duc et al.}{1997}]{1997A&A...326..537D} Duc P.-A., Brinks E., Wink J.~E., Mirabel I.~F., 1997, A\&A, 326, 537

\bibitem[\protect\citeauthoryear{Duc et al.}{2000}]{2000AJ....120.1238D} Duc P.-A., Brinks E., Springel V., Pichardo B., Weilbacher P., Mirabel I.~F., 2000, AJ, 120, 1238. doi:10.1086/301516

\bibitem[\protect\citeauthoryear{Duc \& Mirabel}{1994}]{1994A&A...289...83D} Duc P.-A., Mirabel I.~F., 1994, A\&A, 289, 83

\bibitem[\protect\citeauthoryear{Duc \& Mirabel}{1999}]{1999IAUS..186...61D} Duc, P.-A. \& Mirabel, I.~F.\ 1999, "Tidal Dwarf Galaxies", Proceedings of IAU Symposium Galaxy Interactions at Low and High Redshift, No. 186, pag. 61


\bibitem[\protect\citeauthoryear{Fu et al.}{2020}]{2020arXiv201106368F} Fu Z., Sengupta C., Sethuram R., Pradhan B., Singh M., Misra K., Scott T.~C., et al., 2020, arXiv, arXiv:2011.06368

\bibitem[\protect\citeauthoryear{Gordon, Koribalski, \& Jones}{2001}]{2001MNRAS.326..578G} Gordon S., Koribalski B., Jones K., 2001, MNRAS, 326, 578. doi:10.1046/j.1365-8711.2001.04588.x

\bibitem[\protect\citeauthoryear{Hancock et al.}{2009}]{2009AJ....137.4643H} Hancock M., Smith B.~J., Struck C., Giroux M.~L., Hurlock S., 2009, AJ, 137, 4643. doi:10.1088/0004-6256/137/6/4643

\bibitem[\protect\citeauthoryear{Hibbard et al.}{2005}]{2005ApJ...619L..87H} Hibbard J.~E., Bianchi L., Thilker D.~A., Rich R.~M., Schiminovich D., Xu C.~K., Neff S.~G., et al., 2005, ApJL, 619, L87. doi:10.1086/423244

\bibitem[\protect\citeauthoryear{Hibbard et al.}{2001}]{2001AJ....122.2969H} Hibbard J.~E., van der Hulst J.~M., Barnes J.~E., Rich R.~M., 2001, AJ, 122, 2969. doi:10.1086/324102

\bibitem[\protect\citeauthoryear{Huertas-Company et al.}{2008}]{2008A&A...478..971H} Huertas-Company M., Rouan D., Tasca L., Soucail G., Le F{\`e}vre O., 2008, A\&A, 478, 971

\bibitem[\protect\citeauthoryear{Hunter, Elmegreen \& Ludka}{2010}]{2010AJ....139..447H} Hunter D.~A., Elmegreen B.~G., Ludka B.~C., 2010, AJ, 139, 447

\bibitem[\protect\citeauthoryear{Kaviraj et al.}{2012}]{2012MNRAS.419...70K} Kaviraj S., Darg D., Lintott C., Schawinski K., Silk J., 2012, MNRAS, 419, 70

\bibitem[\protect\citeauthoryear{Kennicutt}{1998}]{1998ARA&A..36..189K} Kennicutt R.~C., 1998, ARA\&A, 36, 189

\bibitem[\protect\citeauthoryear{Lauger et al. }{2005}]{2005A&A...434...77L} Lauger S., Burgarella D., Buat V., 2005, A\&A, 434, 77

\bibitem[\protect\citeauthoryear{Lisenfeld, et al.}{2001}]{2001dge..conf..273L} Lisenfeld U., Braine J., Duc P.-A., Charmandaris V., Vallejo O., Leon S., Brinks E., 2001, dge..conf,  273, dge..conf

\bibitem[\protect\citeauthoryear{Madau, Pozzetti \& Dickinson}{1998}]{1998ApJ...498..106M} Madau P., Pozzetti L., Dickinson M., 1998, ApJ, 498, 106

\bibitem[\protect\citeauthoryear{Maga{\~n}a-Serrano et al.}{2020}]{2020RMxAA..56...39M} Maga{\~n}a-Serrano M.~A., Hidalgo-G{\'a}mez A.~M., Vega-Acevedo I., Casta{\~n}eda H.~O., 2020, RMxAA, 56, 39. doi:10.22201/ia.01851101p.2020.56.01.06

\bibitem[\protect\citeauthoryear{Mirabel, Dottori, \& Lutz}{1992}]{1992A&A...256L..19M} Mirabel I.~F., Dottori H., Lutz D., 1992, A\&A, 256, L19

\bibitem[\protect\citeauthoryear{Lelli et al.}{2015}]{2015A&A...584A.113L} Lelli F., Duc P.-A., Brinks E., Bournaud F., McGaugh S.~S., Lisenfeld U., Weilbacher P.~M., et al., 2015, A\&A, 584, A113. doi:10.1051/0004-6361/201526613

\bibitem[\protect\citeauthoryear{Mayya \& Romano}{2001}]{2001RMxAC..11..115M} Mayya Y.~D., Romano R., 2001, RMxAC, 11, 115

\bibitem[\protect\citeauthoryear{Menanteau et al.}{2006}]{2006AJ....131..208M} Menanteau F., Ford H.~C., Motta V., Ben{\'\i}tez N., Martel A.~R., Blakeslee J.~P., Infante L., 2006, AJ, 131, 208

\bibitem[\protect\citeauthoryear{Meurer et al.}{2009}]{2009ApJ...695..765M} Meurer G.~R., Wong O.~I., Kim J.~H., Hanish D.~J., Heckman T.~M., Werk J., Bland-Hawthorn J., et al., 2009, ApJ, 695, 765. doi:10.1088/0004-637X/695/1/765

\bibitem[\protect\citeauthoryear{Mu{\~n}oz-Elgueta et al.}{2018}]{2018MNRAS.480.3257M} Mu{\~n}oz-Elgueta N., Torres-Flores S., Amram P., Hernandez-Jimenez J.~A., Urrutia-Viscarra F., Mendes de Oliveira C., G{\'o}mez-L{\'o}pez J.~A., 2018, MNRAS, 480, 3257. doi:10.1093/mnras/sty1934

\bibitem[\protect\citeauthoryear{Neichel et al.}{2008}]{2008A&A...484..159N} Neichel B., Hammer F., Puech M., Flores H., Lehnert M., Rawat A., Yang Y., et al., 2008, A\&A, 484, 159. doi:10.1051/0004-6361:20079226

\bibitem[\protect\citeauthoryear{Pearson et al.}{2019}]{2019A&A...631A..51P} Pearson W.~J., Wang L., Alpaslan M., Baldry I., Bilicki M., Brown M.~J.~I., Grootes M.~W., et al., 2019, A\&A, 631, A51

\bibitem[\protect\citeauthoryear{Rosa-Gonz{\'a}lez, Terlevich \& Terlevich}{2002}]{2002MNRAS.332..283R} Rosa-Gonz{\'a}lez D., Terlevich E., Terlevich R., 2002, MNRAS, 332, 283

\bibitem[\protect\citeauthoryear{Sabbi et al.}{2008}]{2008ApJ...676L.113S} Sabbi E., Gallagher J.~S., Smith L.~J., de Mello D.~F., Mountain M., 2008, ApJL, 676, L113. doi:10.1086/587548

\bibitem[\protect\citeauthoryear{Salmon et al.}{2015}]{2015ApJ...799..183S} Salmon B., Papovich C., Finkelstein S.~L., Tilvi V., Finlator K., Behroozi P., Dahlen T., et al., 2015, ApJ, 799, 183. doi:10.1088/0004-637X/799/2/183

\bibitem[\protect\citeauthoryear{Salpeter}{1955}]{1955ApJ...121..161S} Salpeter E.~E., 1955, ApJ, 121, 161

\bibitem[\protect\citeauthoryear{Schweizer}{1978}]{1978IAUS...77..279S} Schweizer, F.\ 1978, "Galaxies with Long Tails", Proceedings of the Symposium Structure and Properties of Nearby Galaxies, No. 77, pag. 279

\bibitem[\protect\citeauthoryear{Scott et al.}{2018}]{2018MNRAS.475.1148S} Scott T.~C., Lagos P., Ramya S., Sengupta C., Paudel S., Sahu D.~K., Misra K., et al., 2018, MNRAS, 475, 1148. doi:10.1093/mnras/stx3248


\bibitem[\protect\citeauthoryear{Sengupta et al.}{2013}]{2013MNRAS.431L...1S} Sengupta C., Dwarakanath K.~S., Saikia D.~J., Scott T.~C., 2013, MNRAS, 431, L1. doi:10.1093/mnrasl/sls039

\bibitem[\protect\citeauthoryear{Sengupta et al.}{2014}]{2014MNRAS.444..558S} Sengupta C., Scott T.~C., Dwarakanath K.~S., Saikia D.~J., Sohn B.~W., 2014, MNRAS, 444, 558. doi:10.1093/mnras/stu1463

\bibitem[\protect\citeauthoryear{Sengupta et al.}{2017}]{2017MNRAS.469.3629S} Sengupta C., Scott T.~C., Paudel S., Dwarakanath K.~S., Saikia D.~J., Sohn B.~W., 2017, MNRAS, 469, 3629. doi:10.1093/mnras/stx885

\bibitem[\protect\citeauthoryear{Schechtman-Rook \& Hess}{2012}]{2012ApJ...750..171S} Schechtman-Rook A., Hess K.~M., 2012, ApJ, 750, 171. doi:10.1088/0004-637X/750/2/171

\bibitem[\protect\citeauthoryear{Smith et al.}{2010}]{2010AJ....139.1212S} Smith B.~J., Giroux M.~L., Struck C., Hancock M., 2010, AJ, 139, 1212. doi:10.1088/0004-6256/139/3/1212

\bibitem[\protect\citeauthoryear{Vega-Acevedo \& Hidalgo-G{\'a}mez}{2014}]{2014ASPC..480..239V} Vega-Acevedo I., Hidalgo-G{\'a}mez A.~M., 2014, ASPC,  239, ASPC..480

\bibitem[\protect\citeauthoryear{Vega-Acevedo}{2013}]{mastersthesisvega} Vega-Acevedo I., 2013, Mc thesis 'Estudio del par\'ametro de asimetr\'ia en galaxias espirales enanas', Escuela Superior de F\'isica y Matem\'aticas, Instituto Polit\'ecnico Nacional, Mexico city


\bibitem[\protect\citeauthoryear{Yagi et al.}{2006}]{2006MNRAS.368..211Y} Yagi M., Nakamura Y., Doi M., Shimasaku K., Okamura S., 2006, MNRAS, 368, 211

\bibitem[\protect\citeauthoryear{Zasov et al.}{2017}]{2017MNRAS.469.4370Z} Zasov A.~V., Saburova A.~S., Egorov O.~V., Uklein R.~I., 2017, MNRAS, 469, 4370. doi:10.1093/mnras/stx1158

\end{thebibliography}
\end{document}